\documentclass[10pt,twocolumn,english,conference,a4paper]{IEEEtran}
\usepackage[T1]{fontenc}
\usepackage[active]{srcltx}
\usepackage{color}
\usepackage{array}
\usepackage{float}
\usepackage{multirow}
\usepackage{amsthm}
\usepackage{amsmath}
\usepackage{graphicx}
\usepackage [latin1]{inputenc}

\makeatletter

\providecommand{\tabularnewline}{\\}
\floatstyle{ruled}
\newfloat{algorithm}{tbp}{loa}
\providecommand{\algorithmname}{Algorithm}
\floatname{algorithm}{\protect\algorithmname}

  \theoremstyle{plain}
  \newtheorem{lem}{\protect\lemmaname}

\ifCLASSOPTIONcompsoc
\usepackage[caption=false,font=normalsize,labelfont=sf,textfont=sf]{subfig}
\else
\usepackage[caption=false,font=footnotesize]{subfig}
\fi

\usepackage{cite}
\usepackage{bm}
\usepackage{algorithmic}
\usepackage{algorithm}
\usepackage{graphicx}
\IEEEoverridecommandlockouts

\makeatother

\usepackage{babel}
\providecommand{\lemmaname}{Lemma}

\begin{document}

\title{\textcolor{black}{Cooperative Beam Hopping for Accurate Positioning
in Ultra-Dense LEO Satellite Networks }}

\author{\IEEEauthorblockN{Yu Wang, Ying Chen, Yunfei Qiao, Hejia~Luo,
Xiaolu Wang, Rong~Li,\textsuperscript{} and Jun Wang \\}\IEEEauthorblockA{\textsuperscript{}Huawei
Technologies Co., Ltd., Hangzhou, China.\\
Email: wangyu207@huawei.com }}
\maketitle
\begin{abstract}
\textcolor{black}{In ultra-dense LEO satellite networks, conventional
communication-oriented beam pattern design cannot provide multiple
favorable signals from different satellites simultaneously, and thus
leads to poor positioning performance. To tackle this issue, in this
paper, we propose a novel cooperative beam hopping (BH) framework
to adaptively tune beam layouts suitable for multi-satellite coordinated
positioning. On this basis, a joint user association, BH design and
power allocation optimization problem is formulated to minimize average
}Cramér-Rao\textcolor{black}{{} lower bound (CRLB). An efficient flexible
BH control algorithm (FBHCA) is then proposed to solve the problem.
Finally, a thorough experimental platform is built following the Third
Generation Partnership Project (3GPP) defined non-terrestrial network
(NTN) simulation parameters to validate the performance gain of the
devised algorithm. The numerical results demonstrate that FBHCA can
significantly improve CRLB performance over the benchmark scheme.}\end{abstract}
\begin{IEEEkeywords}
Multibeam LEO satellite, TDOA positioning, beam hopping, Cramér-Rao\textcolor{black}{{}
lower bound}, 3GPP NTN. \textmd{\normalsize{\vspace{-0.5em}}}{\normalsize \par}
\end{IEEEkeywords}

\section{Introduction\label{sec:introduction} }

\textcolor{black}{ With the advent of New Space era, satellite communication
has gained a renewed upsurge due to its ability to provide global
wireless coverage and continuous service guarantee especially in scenarios
not optimally supported by terrestrial infrastructures\cite{WCM17_ResourceManagement,Survey_SatComintheNewSpaceEra}.
Specifically, some standardization endeavors have been sponsored by
the Third Generation Partnership Project (3GPP) to study a set of
necessary adaptations enabling the operation of 5G New Radio (NR)
protocol in non-terrestrial network (NTN) with a priority on satellite
access \cite{38821_SolutionstoSupportNTN,38811}. In the NTN context,
compared with conventional geostationary earth orbit (GEO) and medium
earth orbit (MEO),  low earth orbit (LEO) based satellite networks
stand out as a promising solution concerning the lower propagation
delay, power consumption and launch cost. As such, numerous companies
have announced ambitious plans to provide broadband Internet access
over the globe by deploying LEO satellite mega-constellations, e.g.,
Oneweb, Kuiper and Starlink. }

\textcolor{black}{On the other hand, the rapid proliferation of location-based
services, e.g., smart transportation, augmented reality and autonomous
driving, bring marvellous value-added opportunities to communication
networks. Besides, location-aided communication optimization such
as enhanced access control and simplified mobility management can
be fully exploited to improve network scalability, latency, and robustness
\cite{LocationAwareCommunication}. Given those benefits, it is imperative
to provide high precision positioning information on top of communication
networks. As stipulated by NR Release-16, a positioning accuracy of
3-meter within 1 second end to end latency should be achieved for
commercial use cases, and subsequent releases are expected to further
realize sub-meter localization accuracy and millisecond level lower
latency \cite{38855_NRPositioning}. This induces great challenge
to meet such stringent requirements.}

\textcolor{black}{To circumvent this issue}, a plenty of research
efforts on localization techniques have been conducted from both the
academic and industry communities \cite{Survey_Localization,Survey_Localization1to5G}.
In terrestrial cellular networks, numerous ranging measurement based
positioning schemes, e.g., time of arrival (TOA), time difference
of arrival (TDOA), angle of arrival (AOA), and received signal strength
(RSS), are devised to figure out source location. Nevertheless, those
positioning schemes cannot be directly applied to LEO satellite communication,
owing to its unique characteristics such as dynamic network topology,
strong multibeam interference, and channel model. While in the satellite
scenario, state of the art positioning studies mainly focus on range
error analysis, e.g., Cramér-Rao lower bound (CRLB), from a macroscopic
constellation design perspective \cite{LEOConstellationforLocalization,Access_MultiSatellitePassiveLocalization}.
However, the concrete multibeam structure design philosophy and underlying
beam management problem to optimize positioning performance remain
unexplored in LEO mega-constellation communication networks.

It is technically challenging to develop accurate positioning schemes
for ultra-dense LEO satellite networks, due to several reasons: 1)
A LEO satellite generally suffers from limited available onboard payload,
and thus only a few physical transceivers/beams can be utilized. These beams
should be efficiently shared for dual functional communication and
positioning purposes. However, conventional communication oriented
multibeam design cannot guarantee multiple strong signals from different
satellites, and thus significantly degrades localization performance;
2) Since there is no near-far effect, the interference problem in
satellite environment becomes critical. Besides, the fast mobility
of LEO satellites makes the multibeam interference more complex and
time-varying \cite{JSAC_Multi-ResourceScheduling}; 3) To support
accurate localization, it is essential to perform coordinate beam
management among multiple satellites. The problem is quite difficult
to handle considering both the nonlinearity of optimization metric
and intrinsic coupling within user association, power/beam allocation
and interference.

In this paper, to address the aforementioned challenges, we investigate
the positioning oriented multibeam pattern design as well as the beam
management problem in LEO mega-constellation communication system.
The main contributions of this paper are summarized as follows.
Firstly, we propose a novel cooperative beam hopping (BH) framework
to flexibly tune the physical beam layout for optimized positioning
usage. On the basis, an average CRLB minimization problem subject
to user association, BH management and power allocation related constraints
is formulated. Secondly, a flexible BH control algorithm (FBHCA) is
proposed to decompose the original problem into three sub-problems,
i.e., user association, BH design and power allocation sub-problems,
which are further solved by max-SINR criteria, Voronoi diagram and
semi-definite programming (SDP) technique, respectively. Finally,
we provide extensive simulations in calibrated 3GPP NTN platform to
evaluate the effectiveness of FBHCA and demonstrate its CRLB performance
gain over existing schemes.

The remainder of this paper is organized as follows. Section \ref{sec:System-Model}
introduces the system model and presents the cooperative BH framework.
The optimization problem formulation and corresponding FBHCA solution
are investigated in Section \ref{sec:Optimization-Problem-Formulation}
and Section \ref{sec:Proposed-Solution}, respectively. Section \ref{sec:Performance-Evaluation}
provides the simulation results, followed by conclusions in Section
\ref{sec:Conclusion}.

\section{System Model and BH Framework \label{sec:System-Model}}

\subsection{System Model}

We consider an orthogonal frequency division multiplexing
(OFDM) based ultra-dense LEO multibeam satellite network. The satellite
system operates at a center frequency $f_{0}$ with total system
bandwidth $W$. To mitigate intra-satellite inter-beam interference,
a frequency/polarization reuse factor of $\rho$ is applied. For inter-satellite
interference reduction, time/frequency/code domain
multiplexing can be utilized, and such kind of interference
is neglected. For example, the positioning reference signal design
with frequency reuse and muting configuration in 3GPP LTE can mitigate
inter-cell interference. In the network, there are a set of $\mathcal{I}=\{1,\ldots,I\}$
LEO satellites. A satellite $i\in\mathcal{I}$ is equipped with a
set of $\mathcal{B}=\{1,\ldots,B\}$ beams. For beam allocation, we
further define a binary variable $\gamma_{i,b}=1$ if beam $i$ is
allocated by satellite $i$, and $\gamma_{i,b}=0$ otherwise. A set
of $\mathcal{J}=\{1,\ldots,J\}$ user equipments (UEs) are distributed
requesting for positioning service. For a user $j\in\mathcal{J}$,
the associated set of satellites used for positioning is denoted as
$\mathcal{I}_{j}=\{1,\ldots,I_{j}\}$. The TDOA positioning
scheme is adopted to calculate UE location results.

As per 3GPP NTN specifications \cite{38811}, the following satellite
antenna pattern is considered

\begin{equation}
G_{t}(\theta)=\begin{cases}
\left(\frac{J_{1}(2\pi f_{0}asin\theta/c)}{\pi f_{0}asin\theta/c}\right)^{2}, & \theta\neq0\\
1, & \theta=0
\end{cases}
\end{equation}
where $J_{1}(\cdot)$ is the Bessel function of the first kind and
first order with the argument, $a$ is the antenna aperture radius,
$\theta$ is the steering angle, and $c$ is the light propagation
speed. Meanwhile, the total path loss consists of following components:\vspace{-1.0em}

\begin{equation}
PL(d)=32.45+20\log_{10}(f_{0}\cdot d)+SF+PL_{g}+PL_{s},
\end{equation}
where $SF$ is the shadow fading modeled by a log-normal distribution
$N(0,\sigma_{SF}^{2})$, $d$ is the slant path distance, and $PL_{g}$
and $PL_{s}$ represent the atmospheric absorption and scintillation
loss, respectively. The power $P_{i,b}^{j}$ received by the $j$-th
user from $b$-th beam of $i$-th satellite is \vspace{-1.0em}

\begin{equation}
P_{i,b}^{j}=10\log_{10}\left(\frac{EIRP_{i,b}\cdot W}{\rho}\right)+G_{t}(\theta_{i,b}^{j})+G_{R}-PL(d_{i,b}^{j}),
\end{equation}
where $\theta_{i,b}^{j}$ and $d_{i,b}^{j}$ denote the angle and
distance between user $j$ and beam $b$ in satellite $i$, respectively.
Besides, $EIRP_{i,b}=P_{i,b}+G_{T}$ is the transmitted Equivalent
Isotropically Radiated Power density allocated to beam $b$ in satellite
$i$, $P_{i,b}$ is the beam transmit power, and $G_{T}$ and $G_{R}$
denote the constant transmit and receive antenna gain, respectively.
To this end, the overall signal-to-interference plus noise ratio (SINR) is \vspace{-1.0em}

\begin{equation}
\beta_{i,b}^{j}=\frac{P_{i,b}^{j}}{\sum_{b^{'}\neq b}P_{i,b^{'}}^{j}+N_{0}},\label{eq:SINR}
\end{equation}
where $N_{0}$ is the noise power determined by UE noise figure and
antenna temperature \cite{38811}.

\subsection{Cooperative Beam Hopping Design}

\textcolor{black}{As a promising solution, BH is devised to employ
only a small subset of transceivers/beams for serving extensive satellite
coverage area. More specifically, assuming that a satellite aims to
achieve a coverage of $B_{cov}$ beams by using $B$ beams with $B<B_{cov}$.
For this purpose, at each time stamp, a maximum number of $B$ beams
are assigned to illuminate a portion of the whole satellite coverage
area, and time-division multiplexing approach is implemented to manipulate
the set of $B$ beams into different portions within the coverage
area of $B_{cov}$ beams. Through flexible beam allocation, full satellite
coverage can be eventually served.}
\begin{figure*}
\centering\includegraphics[scale=1.05]{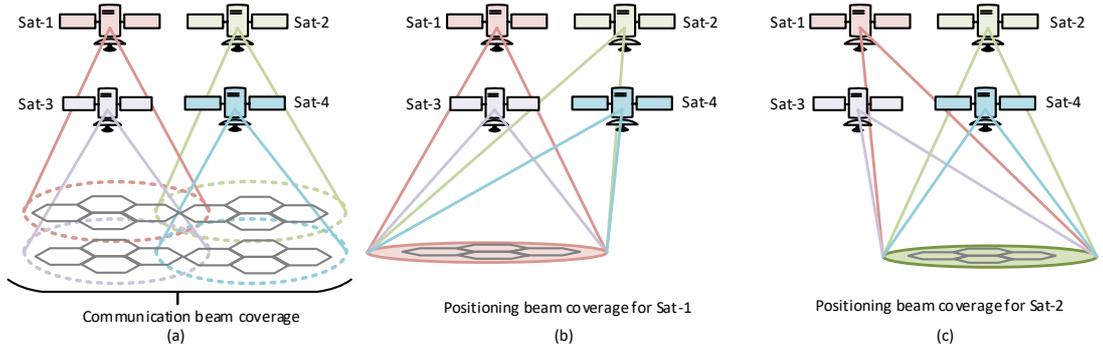}\vspace{-0.5em}

\caption{An example BH system: (a) communication beams; (b) cooperative positioning
beams for Sat-1; (c) cooperative positioning beams for Sat-2.\label{fig:A-beam-hopping} }
\end{figure*}

\textcolor{black}{To support accurate positioning, we exploit BH to
adaptively tune the physical beam layout from communication-oriented
to positioning-oriented design. With cooperative BH from multiple
neighboring satellites, localization performance can be significantly
improved for UEs in a target satellite. An example BH for dual functional
communication and positioning beams in a 4-satellite scenario is depicted
in Fig. \ref{fig:A-beam-hopping}.} On the one hand, communication
beams are generally pointed to coverage area centered at the satellite
nadir as in \textcolor{black}{Fig. \ref{fig:A-beam-hopping}}(a),
such that seamless coverage is formed. On the other hand, to enhance
localization performance, a set of neighbor satellites need to direct
their beams to the target satellite. The center point of positioning
beams normally stays far away from satellite nadir point. The cooperative
BH patterns for positioning of Sat-1 and Sat-2 are sketched in \textcolor{black}{Fig.
\ref{fig:A-beam-hopping}(b) and Fig. \ref{fig:A-beam-hopping}(c),
respectively.}

\begin{figure}
\centering \vspace{-2.5em}\includegraphics[scale=1.0]{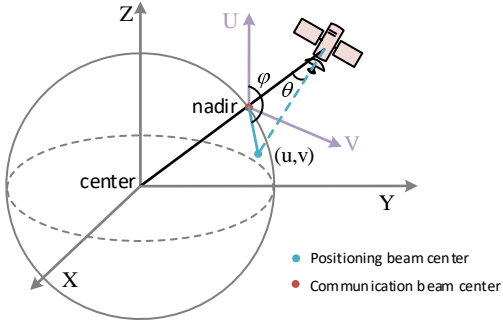}\vspace{-0.5em}

\caption{A sketch of UV plane and satellite geometry.\label{fig:UV plane} } \vspace{-1.5em}
\end{figure}

\textcolor{black}{An efficient UV plane based BH scheme (UVBHS) is
devised for beam operation. As plotted in Fig. \ref{fig:UV plane},
UV plane is defined as the perpendicular plane to the satellite-earth
line on the orbital plane. In UVBHS, a hexagonal beam layout is defined
on the UV plane with UV coordinate of the nadir of the reference satellite
setting to (0,0) for communication beams. For positioning beams, there
are two different configuration situations. Firstly, to perform localization
in its own service area, the satellite can simply reuse the communication
configurations for positioning usage. Secondly, to assist localization
for a neighboring satellite, the satellite needs to translate the
beams centered at (0,0) to the center of the neighbor satellite's
nadir in the UV plane denoted by $(u,v)$. We can derive $u=sin\theta cos\varphi$
and $v=sin\theta sin\varphi$, where $\theta$ and $\varphi$ represent
beam bore-sight steering angle and azimuth, respectively.}

\section{Optimization Problem Formulation\label{sec:Optimization-Problem-Formulation}}

\subsection{CRLB Analysis}

\textcolor{black}{In the mean squared error (MSE) sense, CRLB gives
the lowest possible variance that an unbiased linear estimator can
achieve.} To this end, we use CRLB as the accuracy indicator for TDOA
positioning error analysis.

\subsubsection{TOA Measurement in OFDM system}

For TDOA positioning, a set of TOA measurements from multiple satellites
should be conducted at first. In an OFDM system, denote $S_{l}(k)$
as the signal allocated on the $k$-th subcarrier of the $l$-th symbol.
The number of symbols and subcarriers used for positioning is expressed
as $N_{s}$ and $K$, respectively. According to \cite{CRLBToA},
under a static AWGN channel, the MSE $\sigma_{i}$ of TOA measurement
$\tau_{i}$ at satellite $i$ is calculated by \vspace{-0.5em}

\begin{equation}
\sigma_{i}^{2}\geq CRLB(\tau_{i})=\frac{T_{s}^{2}}{8\pi^{2}\beta_{i,b}^{j}\delta_{i,b}^{j}\Gamma},
\end{equation}
where $T_{s}$ is the OFDM symbol duration, $\delta_{i,b}^{j}$ denotes
the association variable of user $j$ to beam $b$ in satellite $i$,
and $\Gamma=\sum_{l=0}^{N_{s}-1}\sum_{k=-K/2}^{K/2-1}k^{2}|S_{l}(k)|^{2}$.

\subsubsection{CRLB analysis for TDOA positioning}

Based on the above TOA measurements, the CRLB for TDOA positioning
at a target UE is explicitly analyzed herein. The positions of UE
$j\in\mathcal{J}$ and satellites $i\in\mathcal{I}_{j}$ are denoted
as $\boldsymbol{s}=(x,y,z)^{T}$ and $\boldsymbol{s_{i}}=(x_{i},y_{i},z_{i})^{T},i=1,\ldots,I_{j}$,
respectively. The distance between the target UE $j$ and satellite
$i$ is expressed as \vspace{-0.5em}

\begin{equation}
d_{i}=||\boldsymbol{s}-\boldsymbol{s_{i}}||=\sqrt{(\boldsymbol{s}-\boldsymbol{s_{i}})^{T}(\boldsymbol{s}-\boldsymbol{s_{i}})}.\label{eq:range}
\end{equation}

Without loss of generality, we choose $\boldsymbol{s_{1}}$ as the
reference satellite location. Under a line of sight (LOS) scenario
prevalent in satellite context, the TDOA received by $\boldsymbol{s_{i}}$
and $\boldsymbol{s_{1}}$ is \vspace{-1.5em}

\begin{equation}
\tau_{i1}\equiv\tau_{i}-\tau_{1}=\frac{1}{c}(d_{i}-d_{1})+(e_{i}-e_{1}),\label{eq:time}
\end{equation}
where $e_{i}$ is the TOA measurement noise with zero mean and covariance
$\sigma_{i}^{2}$. The CRLB for estimating $\boldsymbol{s}$ in user
$j$ equals\vspace{-0.5em}

\begin{equation}
CRLB_{j}(\boldsymbol{s})=\sqrt{\textrm{tr}\{(\boldsymbol{A}^{T}\boldsymbol{R}^{-1}\boldsymbol{A})^{-1}\}},\label{eq:CRLB}
\end{equation}
where $\boldsymbol{A}=\frac{1}{c}[\frac{(\boldsymbol{s}-\boldsymbol{s_{2}})^{T}}{d_{2}}-\frac{(\boldsymbol{s}-\boldsymbol{s_{1}})^{T}}{d_{1}},\ldots,\frac{(\boldsymbol{s}-\boldsymbol{s_{I_{j}}})^{T}}{d_{I_{j}}}-\frac{(\boldsymbol{s}-\boldsymbol{s_{1}})^{T}}{d_{1}}]^{T}$,
$\textrm{tr}\{\cdot\}$ is the trace operator of matrix, and

\[
\boldsymbol{R}=\left[\begin{array}{ccc}
\sigma_{1}^{2}+\sigma_{2}^{2} & \cdots & \sigma_{1}^{2}\\
\vdots & \ddots & \vdots\\
\sigma_{1}^{2} & \cdots & \sigma_{1}^{2}+\sigma_{I_{j}}^{2}
\end{array}\right].
\]

\subsection{Average CRLB Minimization Problem Optimization}

\textcolor{black}{Based on the envisioned cooperative BH framework,
i.e., UVBHS, an average CRLB optimization problem is formulated to
improve positioning accuracy.}\vspace{-1.0em}

\begin{eqnarray}
 & \mathbf{} & \min_{\boldsymbol{\delta},\boldsymbol{\gamma},\boldsymbol{P}}\:\frac{1}{J}\sum_{j\in\mathcal{J}}CRLB_{j}(\boldsymbol{s})\nonumber \\
 &  & \textrm{s.t.}\:\:\:\textrm{C1}:\:\sum_{b\in\mathcal{B}}\delta_{i,b}^{j}\leq1,\:\forall j\in\mathcal{J},\: i\in\mathcal{I},\: b\in\mathcal{B}\nonumber \\
 &  & \:\:\:\:\:\:\:\:\textrm{C2}:\:\sum_{j\in\mathcal{J}}\delta_{i,b}^{j}\leq\Lambda\cdot\gamma_{i,b},\:\forall i\in\mathcal{I},\: b\in\mathcal{B}\nonumber \\
 &  & \:\:\:\:\:\:\:\:\textrm{C3}:\: P_{i,b}\leq P_{tot}^{beam},\:\forall i\in\mathcal{I},\: b\in\mathcal{B}\nonumber \\
 &  & \:\:\:\:\:\:\:\:\textrm{C4}:\:\sum_{b\in\mathcal{B}}\gamma_{i,b}P_{i,b}\leq P_{tot}^{sat},\:\forall i\in\mathcal{I}\nonumber \\
 &  & \:\:\:\:\:\:\:\:\textrm{C5}:\:\delta_{i,b}^{j}\in\{0,1\},\:\forall j\in\mathcal{J},\: i\in\mathcal{I}\nonumber \\
 &  & \:\:\:\:\:\:\:\:\textrm{C6}:\:\gamma_{i,b}\in\{0,1\},\:\forall i\in\mathcal{I},\: b\in\mathcal{B}.\label{eq:Opt_Problem}
\end{eqnarray}

In problem (\ref{eq:Opt_Problem}), the objective is to minimize average
CRLB of UEs by optimizing decision variables $\boldsymbol{\delta}=\{\delta_{i,b}^{j}\}$,
$\boldsymbol{P}=\{P_{i,b}\}$, and $\boldsymbol{\gamma}=\{\gamma_{i,b}\}$.
Constraints C1 states that each user should be associated to at most
one beam at a given satellite; C2 specifies that the beam should be
allocated once a user is associated to the beam, where $\Lambda$
is a sufficiently large integer; C3 is the individual beam power constraint;
and C4 is imposed to guarantee that the allocated power at a satellite
should not exceed its total available power. The NP-hardness of the
optimization problem is shown below.
\begin{lem}
Problem (\ref{eq:Opt_Problem}) is NP-hard. \end{lem}
\begin{IEEEproof}
Consider a generalized case where power related constraints C3 and
C4 are relaxed. To this end, (\ref{eq:Opt_Problem}) is reduced to
the sensor selection problem\cite{SensorSelectionProblem}, which
is NP-hard with nonlinear objective and non-convex constraints. Thus,
the NP-hardness of the original problem is derived as well.
\end{IEEEproof}

\section{Proposed Solution\label{sec:Proposed-Solution}}

Herein, the problem is first decomposed and an efficient FBHCA solution
is proposed, followed by complexity analysis. \vspace{-1.0em}

\subsection{Problem Decomposition based Solution }

The problem (\ref{eq:Opt_Problem}) is solved by decomposing it into
three sub-problems, namely, user association sub-problem, BH design
sub-problem, and power allocation sub-problem.

\subsubsection{User association}

To solve the user association sub-problem, the following lemma is
introduced.
\begin{lem}
\label{lemma:CRLB}The $CRLB_{j}(\boldsymbol{s})$ is a monotonically
decreasing function of \textup{$\beta_{i,b}^{j}$.}\end{lem}
\begin{IEEEproof}
Define $N=I_{j}$ and $\boldsymbol{R}=\boldsymbol{R}_{0}+\mu\nu^{T}$,
where $\mu=\nu=(\sigma_{1},\ldots,\sigma_{1})^{T}$ are column vectors
with a length of $N-1$. Besides, $\boldsymbol{R}_{0}=\textrm{diag}\{\sigma_{2}^{2},\ldots,\sigma_{N}^{2}\}$.
According to the Matrix Inversion Lemma, we obtain\vspace{-0.5em}
\begin{eqnarray}
 & \boldsymbol{R}^{-1} & =(\boldsymbol{R}_{0}+\mu\nu^{T})^{-1}\nonumber \\
 &  & =\boldsymbol{R}_{0}^{-1}-\frac{\boldsymbol{R}_{0}^{-1}\mu\nu^{T}\boldsymbol{R}_{0}^{-1}}{1+\nu^{T}\boldsymbol{R}_{0}^{-1}\mu}\nonumber \\
 &  & =\boldsymbol{R}_{0}^{-1}-\left[\begin{array}{ccc}
\frac{1}{\sigma_{2}^{2}\sigma_{2}^{2}\Omega} & \cdots & \frac{1}{\sigma_{2}^{2}\sigma_{N}^{2}\Omega}\\
\vdots & \ddots & \vdots\\
\frac{1}{\sigma_{N}^{2}\sigma_{2}^{2}\Omega} & \cdots & \frac{1}{\sigma_{N}^{2}\sigma_{N}^{2}\Omega}
\end{array}\right]\label{eq:R_inv}\\
 &  & \equiv\boldsymbol{R}_{0}^{-1}-\boldsymbol{H},\nonumber
\end{eqnarray}
where $\Omega=\frac{1}{\sigma_{1}^{2}}+\frac{1}{\sigma_{2}^{2}}+\cdots+\frac{1}{\sigma_{N}^{2}}$.
On the basis, the CRLB expression can be equivalently written as follows\vspace{-0.5em}
\begin{eqnarray}
 &  & \textrm{tr}\{(\boldsymbol{A}^{T}\boldsymbol{R}^{-1}\boldsymbol{A})^{-1}\}=\textrm{tr}\left\{ \left(\boldsymbol{A}^{T}\left(\boldsymbol{R}_{0}^{-1}-\boldsymbol{H}\right)\boldsymbol{A}\right)^{-1}\right\} \nonumber \\
 &  & =\textrm{tr}\left\{ \left(\boldsymbol{A}^{T}\boldsymbol{R}_{0}^{-1}\boldsymbol{A}-\boldsymbol{A}^{T}\boldsymbol{H}\boldsymbol{A}\right)^{-1}\right\} \label{eq:CRLB_ref}\\
 &  & =\textrm{tr}\left\{ \boldsymbol{Y}^{-1}\right\} +\textrm{tr}\left\{ \boldsymbol{Y}^{-1}\boldsymbol{Z}\left(\boldsymbol{Z}-\boldsymbol{Z}\boldsymbol{Y}^{-1}\boldsymbol{Z}\right)^{-1}\boldsymbol{Z}\boldsymbol{Y}^{-1}\right\} ,\nonumber
\end{eqnarray}
where $\boldsymbol{Y}=\boldsymbol{A}^{T}\boldsymbol{R}_{0}^{-1}\boldsymbol{A}$,
and $\boldsymbol{Z}=\boldsymbol{A}^{T}\boldsymbol{H}\boldsymbol{A}$.
Consider that $0<\frac{1}{\sigma_{i}^{2}\Omega}\ll1$ holds for a
typical LEO satellite network, and thus the value of $\textrm{tr}\left\{ \boldsymbol{Y}^{-1}\right\} $
dominates the final result. Consequently, $CRLB_{j}(\boldsymbol{s})$
decreases as $\beta_{i,b}^{j}$ increases.
\end{IEEEproof}
According to Lemma \ref{lemma:CRLB}, if the initial power allocation
$\boldsymbol{P}^{(0)}$ is given, the SINR $\beta_{i,b}^{j}$ can
be computed using (\ref{eq:SINR}). Therefore, $\delta_{i,b}^{j}$
is optimally solved by attaching user $j$ to the beam with maximum
SINR for satellite $i$, which is \vspace{-0.5em}

\begin{equation}
\delta_{i,b}^{j}=\begin{cases}
1, & \beta_{i,b}^{j}=\max_{b\in\mathcal{B}}\left\{ \beta_{i,b}^{j}\right\} \\
0, & \textrm{otherwise}.
\end{cases}\label{eq:user_association}
\end{equation}
Note that for positioning purpose, association to multiple satellites
is required for a UE. This is different from user attachment to only
one serving satellite in communication oriented design. Given maximum
SINR based user association criteria, the decisions of $\boldsymbol{\delta}=\{\delta_{i,b}^{j}\}$
are easily obtained.

\subsubsection{BH design }

\begin{figure}
\centering\vspace{-1.0em}\includegraphics[scale=0.55]{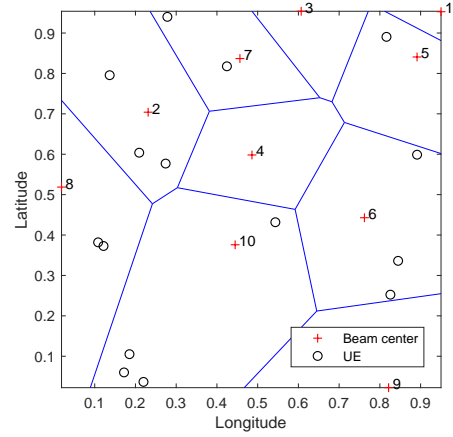}\vspace{-0.5em}

\caption{An example of Voronoi diagram for a 10-beam satellite.\label{fig:Voronoi} } \vspace{-1.5em}
\end{figure}

After solving $\boldsymbol{\delta}=\{\delta_{i,b}^{j}\}$, problem
(\ref{eq:Opt_Problem}) can be further reformulated as \vspace{-1.0em}

\begin{eqnarray}
 & \mathbf{} & \min_{\boldsymbol{\gamma},\boldsymbol{P}}\:\frac{1}{J}\sum_{j\in\mathcal{J}}\left(\textrm{tr}\left\{ \boldsymbol{Y}^{-1}\right\} +\textrm{tr}\left\{ \boldsymbol{X}\right\} \right)\nonumber \\
 &  & \textrm{s.t.}\:\:\:\textrm{C2},\textrm{C3},\textrm{C4},\textrm{C6}\nonumber \\
 &  & \:\:\:\:\:\:\:\:\textrm{C7}:\:\left[\begin{array}{cc}
\boldsymbol{Z}-\boldsymbol{Z}\boldsymbol{Y}^{-1}\boldsymbol{Z} & \boldsymbol{Z}\boldsymbol{Y}^{-1}\\
\boldsymbol{Y}^{-1}\boldsymbol{Z} & \boldsymbol{X}
\end{array}\right]\succeq0,\label{eq:Opt_Problem_Reformulation}
\end{eqnarray}
where the objective is rewritten by using (\ref{eq:CRLB_ref}), and
the auxiliary variable $\boldsymbol{X}$ satisfies \vspace{-0.5em}

\begin{equation}
\boldsymbol{Y}^{-1}\boldsymbol{Z}\left(\boldsymbol{Z}-\boldsymbol{Z}\boldsymbol{Y}^{-1}\boldsymbol{Z}\right)^{-1}\boldsymbol{Z}\boldsymbol{Y}^{-1}\preceq\boldsymbol{X}.
\end{equation}
Meanwhile, the constraint C7 in reformulated problem (\ref{eq:Opt_Problem_Reformulation})
is added based on the Schur complement law \cite{SensorSelectionProblem}.
Since the only non-convex decision variables in problem (\ref{eq:Opt_Problem_Reformulation})
are $\boldsymbol{\gamma}=\{\gamma_{i,b}\}$, an efficient Voronoi
diagram \cite{LEO3DVoronoiDiagram} is used to solve the beam allocation
problem. More specifically, each beam corresponds to a polygon representing
its serving coverage area. We set $\gamma_{i,b}=1$ if at least one
user lies in the Voronoi area, and $\gamma_{i,b}=0$ otherwise. To
this end, the BH design sub-problem is successfully tackled. An example
Voronoi graph for a 10-beam satellite is given in Fig. \ref{fig:Voronoi},
where a polygon formed by several blue lines corresponds to a beam
serving area. Note that because no UE attaches to beam 4 in Fig. \ref{fig:Voronoi},
$\gamma_{i,4}=0$ can be determined for the satellite.

\subsubsection{Power allocation}

Until now, the original problem is reduced to a power allocation problem
with convex constraints\vspace{-1.5em}

\begin{eqnarray}
 & \mathbf{} & \min_{\boldsymbol{P}}\:\frac{1}{J}\sum_{j\in\mathcal{J}}\left(\textrm{tr}\left\{ \boldsymbol{Y}^{-1}\right\} +\textrm{tr}\left\{ \boldsymbol{X}\right\} \right)\nonumber \\
 &  & \textrm{s.t.}\:\:\:\textrm{C3},\textrm{C4},\textrm{C7}.\label{eq:PowerAllocation}
\end{eqnarray}
The above sub-problem (\ref{eq:PowerAllocation}) is a typical SDP
problem, and can be efficiently solved by existing convex optimization
methods, e.g., the interior-point method \cite{Book_ConvexOpt}. The
obtained power results are capitalized to update the preceding power
vector. Then, the algorithm moves forward to the next iteration. The
iterations cease until a predetermined number $M$ is reached. The
detailed procedure of FBHCA is summarized in Algorithm \ref{alg:FBHCA}.
\begin{algorithm}[t]
\caption{FBHCA\label{alg:FBHCA}}

\begin{algorithmic}[1]

\STATE \textbf{Initialization:} Set iteration number $M$, initialize
$m=0$, and $\boldsymbol{P}^{(0)}=\{P_{i,b}\}$, where $P_{i,b}=P_{tot}^{beam}/B$.

\WHILE {$m<M$}

\STATE\textbf{User association:} Compute SINR $\beta_{i,b}^{j}$
and obtain $\delta_{i,b}^{j}$ using maximum SINR criteria in (\ref{eq:user_association});

\STATE \textbf{BH design:} Construct Voronoi graph for beams in each
satellite involved in the positioning process;

\IF {At least one user lies in the Voronoi area}

\STATE Set $\gamma_{i,b}=1$;

\ELSE

\STATE Set $\gamma_{i,b}=0$;

\ENDIF

\STATE \textbf{Power allocation:} Solve (\ref{eq:PowerAllocation})
with the interior-point method and calculate $CRLB^{(m)}(\boldsymbol{s})$;

\STATE \textbf{$m=m+1$};

\ENDWHILE

\STATE Return the final solution with minimum $CRLB^{(m)}(\boldsymbol{s})$.

\end{algorithmic}
\end{algorithm}

\subsection{Complexity Analysis }

The computational complexity of the devised algorithm is discussed
herein. Recall that there are $J$ users, $B$ beams per satellite,
and $N=I_{j}$ satellites used for positioning. The complexity of
FBHCA comprises three parts: 1) User association with the complexity
of $O(JNB)$. The complexity comes from a total of $JNB$ SINR calculations
and comparisons for all UEs; 2) BH design with the complexity of $O(JNB^{2})$.
This is because for a $B$ points Voronoi diagram, the complexity
is $O(B^{2})$ \cite{VoronoiComplexity}. Besides, the Voronoi graph
is generated for $N$ satellites, and $J$ calculations are required
to decide the beam allocation variable $\gamma_{i,b}$. This results
in a total complexity of $O(JNB^{2})$; 3) Power allocation with the
complexity of $O(NB^{4.5})$. For each satellite, solving the SDP
problem incurs $O(B^{4.5})$ worst-case complexity, and a set of $N$
satellites needs to be calculated for power allocation.

The algorithm runs iteratively to obtain the desired solution. There
are $M$ iterations in total, with each iteration generating a complexity
of $O(JNB)+O(JNB^{2})+O(NB^{4.5})=O(JNB^{2}+NB^{4.5})$. Overall,
the computational complexity of FBHCA is derived as $O\left(M(JNB^{2}+NB^{4.5})\right)$.

\section{Performance Evaluation\label{sec:Performance-Evaluation}}

In this section, simulation settings following 3GPP NTN assumptions
are first configured. Afterwards, numerical results are presented
to verify the effectiveness of the algorithm.\vspace{-0.5em}

\subsection{Simulation Settings}

 The set of key parameters used for simulations are summarized in
Table \ref{tab:Key-simulation-paramters.}. Particularly, the satellite
network comprises a total of 2400 satellites, with an orbit height
varying from 800 km to 1500 km and inclination of 87.5 degree. The
synchronization signal blocks (SSBs) in 3GPP NR are used for ToA measurements.
The value of $\sigma_{SF}^{2}$ for shadow fading is a function of
elevation angle following Table 6.6.2-3 in \cite{38811}. For dynamic
simulation, the orbit period is divided into 100 snapshots of equal
time duration. A total of 500 stationary UEs are randomly deployed
in the target area with longitude and latitude setting to {[}-70,-60{]}
and {[}-5,5{]}, respectively.\textcolor{black}{{} Note that all simulations
are performed by extending the already calibrated platform for 3GPP
NTN \cite{HW_Tdoc}. }
\begin{table}[t]
\textcolor{black}{\caption{\textcolor{black}{Key simulation parameters\label{tab:Key-simulation-paramters.}. }}
}\centering\textcolor{black}{}%
\begin{tabular}{|l|l|}
\hline
Parameters  & \textcolor{black}{Values}\tabularnewline
\hline
\textcolor{black}{The number of orbit} & \textcolor{black}{40}\tabularnewline
\hline
The number of satellite per orbit & 60\tabularnewline
\hline
Orbit inclination & \textcolor{black}{87.5}\tabularnewline
\hline
\textcolor{black}{Orbit height} & {[}800,1500{]} km\tabularnewline
\hline
The number of beam per satellite  & 61\tabularnewline
\hline
Carrier frequency & 2 GHz\tabularnewline
\hline
Satellite transmit antenna gain  & 30 dBi\tabularnewline
\hline
Maximum beam transmit power  & 110 W\tabularnewline
\hline
Total satellite power  & 6100 W\tabularnewline
\hline
System bandwidth & 30 MHz\tabularnewline
\hline
Frequency reuse factor & 3\tabularnewline
\hline
Equivalent satellite antenna aperture & 0.5 m\tabularnewline
\hline
Channel model & Clear sky with LOS\tabularnewline
\hline
Atmospheric absorption loss & 0.1 dB\tabularnewline
\hline
Scintillation loss & 2.2 dB\tabularnewline
\hline
UE noise figure  & 7 dB\tabularnewline
\hline
Antenna temperature & 290 K\tabularnewline
\hline
Subcarrier spacing & 15 KHz\tabularnewline
\hline
Number of satellites for positioning & 4, 6, 8\tabularnewline
\hline
\end{tabular}
\end{table}

\subsection{Numerical Results}\vspace{-0.5em}

\begin{figure}[t]
\centering\includegraphics[scale=0.55]{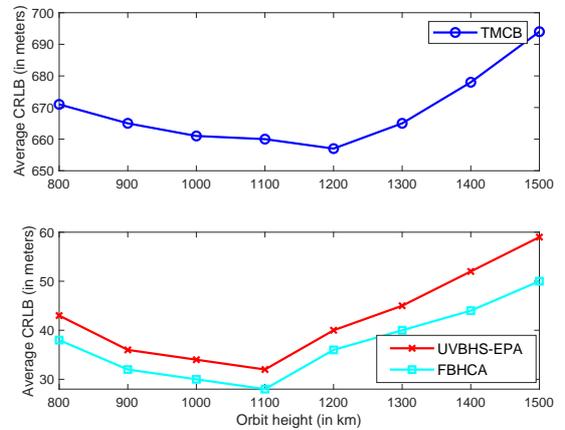}\vspace{-0.5em}

\caption{Average CRLB performance of different schemes versus orbital height
with 4 positioning satellites.\label{fig:CRLB_orbit} } \vspace{-1.5em}
\end{figure}
\textcolor{black}{}
\begin{table}[t]
\textcolor{black}{\caption{\textcolor{black}{SNR values for different schemes\label{tab:SINR_value}. }}
}\centering\textcolor{black}{}%
\begin{tabular}{|l|l|l|l|l|}
\hline
\multirow{2}{*}{Algorithms} & \multicolumn{4}{c|}{SNR for different satellites}\tabularnewline
\cline{2-5}
 & \textcolor{black}{Satellite 1} & Satellite 2 & Satellite 3 & Satellite 4\tabularnewline
\hline
\textcolor{black}{TMCB} & 14.2 & -6.4 & -12.0 & -10.9\tabularnewline
\hline
UVBHS+EPA & 14.2 & 10.2 & 9.8 & 13.9\tabularnewline
\hline
FBHCA & 14.6 & 10.7 & 10.4 & 14.0\tabularnewline
\hline
\end{tabular}
\end{table}\vspace{-0.5em}

For comparison, two benchmark algorithms are used. One is the \textcolor{black}{traditional
method using communication beams (TMCB) for multi-satellite positioning
as shown in Fig. \ref{fig:A-beam-hopping}}(a)\textcolor{black}{.
In TMCB, maximum SINR based user association and equal power allocation
among all satellite beams are assumed. The other one is termed as
UVBHS-EPA, which combines the proposed UVBHS framework with equal
power allocation.}

\textcolor{black}{Fig. \ref{fig:CRLB_orbit} depicts the average CRLB
performance for different schemes as orbital height changes. As orbit
height increases, the CRLB of all the three algorithms first decreases.
The reason is that the constellation of 2400 satellites is not sufficient
to provide full coverage for lower orbit height than around 1100 km.
Lower orbit can cause worse SINR and positioning performance. With
orbit height further rising, the CRLB increases as well due to the
larger experienced path loss. Besides, both UVBHS-EPA and FBHCA significantly
outperform TMCB in terms of CRLB. This is because in TMCB, although
the received signal quality from the serving satellite is favorable,
the signal quality from neighboring satellites is very poor due to
the long distance between UE and beam center. Nonetheless, in the
other two methods, neighboring satellite beams are directed to cover
UEs of interest, and thus multiple signals with good quality can be
measured to enhance positioning accuracy.} To verify the performance
gain, we list the SNR values of different satellites in Table \ref{tab:SINR_value}.

\begin{figure}[t]
\centering\vspace{-1.0em}\includegraphics[scale=0.55]{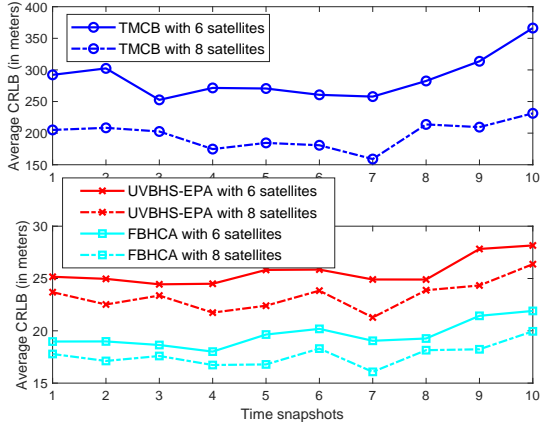}\vspace{-0.5em}

\caption{Average CRLB performance versus time snapshots.\label{fig:CRLB} } \vspace{-1.5em}
\end{figure}

\textcolor{black}{The CRLB results for different schemes as time snapshot
varies are given in Fig \ref{fig:CRLB}. The orbit height is fixed
to 1200 km in the simulation. All the algorithms exhibit CRLB fluctuation
as time snapshots change. The variation tendency is quite complicated,
due to the intertwined influence by time varying inter-beam interference
and dynamic geometric dilution of precision. Besides, as the number
of positioning satellites increases from 6 to 8, the CRLBs for all
the three algorithms improve as well. This phenomena can be expected,
because more signals and better network geometry are obtained by exploring
satellite diversity. }

\section{Conclusion\label{sec:Conclusion}}

In this paper, we have investigated the problem of high accuracy positioning
in ultra-dense LEO satellite networks. A BH framework is proposed
for flexible beam operation. With the framework, we further devise
an efficient FBHCA solution to handle the joint user association,
BH design, and power allocation problem. Significant positioning performance
improvement in terms of average CRLB is demonstrated through 3GPP
NTN simulation platform. As for future work, we tend to take inter-satellite
interference issue into account and deal with the positioning reference
signal design problem to further enhance localization accuracy.

\bibliographystyle{IEEEtran}
\bibliography{ref_NTN_HW}

\end{document}